\documentclass[10pt,letterpaper]{article}
\usepackage{opex3}

\usepackage{amsmath}
\usepackage{amssymb}
\usepackage{cite}

\usepackage[colorlinks=true,
            pdfstartview=FitV, 
            linkcolor=blue, 
            citecolor=blue,
            urlcolor=blue]{hyperref}

\begin{document}

\title{Saturable Lorentz model for fully explicit three-dimensional modeling of nonlinear optics}

\author{Charles Varin,$^*$ Graeme Bart, Rhys Emms, and Thomas Brabec}
\address{Center for Research in Photonics, University of Ottawa, Ottawa (ON) K1N 6N5, Canada}
\email{*cvarin@uottawa.ca}

\begin{abstract}
Inclusion of the instantaneous Kerr nonlinearity in the FDTD framework leads to implicit equations that have to be solved iteratively. In principle, explicit integration can be achieved with the use of anharmonic oscillator equations, but it tends to be unstable and inappropriate for studying strong-field phenomena like laser filamentation. In this paper, we show that nonlinear susceptibility can be provided instead by a harmonic oscillator driven by a nonlinear force, chosen in a way to reproduce the polarization obtained from the solution of the quantum mechanical two-level equations. The resulting saturable, nonlinearly-driven, harmonic oscillator model reproduces quantitatively the quantum mechanical solutions of harmonic generation in the under-resonant limit, up to the 9th harmonic. Finally, we demonstrate that fully explicit leapfrog integration of the saturable harmonic oscillator is stable, even for the intense laser fields that characterize laser filamentation and high harmonic generation. 
\end{abstract}

\ocis{(190.0190) Nonlinear optics; (270.6620) Strong-field processes; (000.3860) Mathematical methods in physics; (000.4430) Numerical approximation and analysis.}

\bibliographystyle{osajnl}
\bibliography{bibfile}

\begin{thebibliography}{10}
\newcommand{\enquote}[1]{``#1''}

\bibitem{taflove2005computational}
A.~Taflove and S.~Hagness, \emph{Computational Electrodynamics: The
  Finite-difference Time-domain Method} (Artech House, 2005).

\bibitem{Varin2014}
C.~Varin, C.~Peltz, T.~Brabec, and T.~Fennel, \enquote{{Light wave driven
  electron dynamics in clusters},} Ann. Phys. \textbf{526}, 135--156 (2014).

\bibitem{boyd2008nonlinear}
R.~Boyd, \emph{Nonlinear Optics, 3rd ed.} (Elsevier Science, 2008).

\bibitem{Varin2012}
C.~Varin, C.~Peltz, T.~Brabec, and T.~Fennel, \enquote{{Attosecond plasma wave
  dynamics in laser-driven cluster nanoplasmas},} Phys. Rev. Lett.
  \textbf{108}, 175007 (2012).

\bibitem{Peltz2012}
C.~Peltz, C.~Varin, T.~Brabec, and T.~Fennel, \enquote{{Fully microscopic
  analysis of laser-driven finite plasmas using the example of clusters},} New
  J. Phys. \textbf{14}, 065011 (2012).

\bibitem{Peltz2014}
C.~Peltz, C.~Varin, T.~Brabec, and T.~Fennel, \enquote{{Time-resolved x-ray
  imaging of anisotropic nanoplasma expansion},} Phys. Rev. Lett. \textbf{113},
  133401 (2014).

\bibitem{Goorjian1992}
P.~Goorjian, A.~Taflove, R.~Joseph, and S.~Hagness, \enquote{{Computational
  modeling of femtosecond optical solitons from Maxwell's equations},} IEEE J.
  Quantum Electron. \textbf{28}, 2416--2422 (1992).

\bibitem{Greene2006}
J.~H. Greene and A.~Taflove, \enquote{{General vector auxiliary differential
  equation finite-difference time-domain method for nonlinear optics},} Opt.
  Express \textbf{14}, 8305 (2006).

\bibitem{taflove2013advances}
A.~Taflove, A.~Oskooi, and S.~Johnson, \emph{Advances in FDTD Computational
  Electrodynamics: Photonics and Nanotechnology} (Artech House, 2013).

\bibitem{Berge2007}
L.~Berg\'{e}, S.~Skupin, R.~Nuter, J.~Kasparian, and J.-P. Wolf,
  \enquote{{Ultrashort filaments of light in weakly ionized, optically
  transparent media},} Reports Prog. Phys. \textbf{70}, 1633--1713 (2007).

\bibitem{couaironPR2007}
A.~Couairon and A.~Mysyrowicz, \enquote{{Femtosecond filamentation in
  transparent media},} Phys. Rep. \textbf{441}, 47--189 (2007).

\bibitem{yee1966}
K.~Yee, \enquote{Numerical solution of initial boundary value problems
  involving {Maxwell's} equations in isotropic media,} IEEE Trans. Antennas
  Propag. \textbf{14}, 302--307 (1966).

\bibitem{stratton1941electromagnetic}
J.~Stratton, \emph{Electromagnetic Theory} (McGraw-Hill, 1941).

\bibitem{jackson1998electromagnetic}
J.~D. Jackson, \emph{{Classical Electrodynamics, 3rd ed.}} (Wiley, 1998).

\bibitem{Gordon2013}
D.~Gordon, M.~Helle, and J.~Pe\~{n}ano, \enquote{{Fully explicit nonlinear
  optics model in a particle-in-cell framework},} J. Comput. Phys.
  \textbf{250}, 388--402 (2013).

\bibitem{allen1975optical}
L.~Allen and J.~Eberly, \emph{Optical Resonance and Two-level Atoms} (Dover,
  1975).

\bibitem{Volkova2011}
E.~A. Volkova, A.~M. Popov, and O.~V. Tikhonova, \enquote{{Nonlinear
  polarization response of an atomic gas medium in the field of a
  high-intensity femtosecond laser pulse},} JETP Lett. \textbf{94}, 519--524
  (2011).

\bibitem{Kohler2013}
C.~K\"{o}hler, R.~Guichard, E.~Lorin, S.~Chelkowski, A.~D. Bandrauk,
  L.~Berg\'{e}, and S.~Skupin, \enquote{{Saturation of the nonlinear refractive
  index in atomic gases},} Phys. Rev. A \textbf{87}, 043811 (2013).

\bibitem{Spott2014}
A.~Spott, A.~Jaroń-Becker, and A.~Becker, \enquote{{Ab initio and perturbative
  calculations of the electric susceptibility of atomic hydrogen},} Phys. Rev.
  A \textbf{90}, 013426 (2014).

\bibitem{Ciddor96}
P.~E. Ciddor, \enquote{{Refractive index of air: new equations for the visible
  and near infrared},} Appl. Opt. \textbf{35}, 1566--1573 (1996).

\bibitem{corkumPRL1993}
P.~B. Corkum, \enquote{Plasma perspective on strong field multiphoton
  ionization,} Phys. Rev. Lett. \textbf{71}, 1994--1997 (1993).

\bibitem{LewensteinPRA1994}
M.~Lewenstein, P.~Balcou, M.~Y. Ivanov, A.~L'Huillier, and P.~B. Corkum,
  \enquote{Theory of high-harmonic generation by low-frequency laser fields,}
  Phys. Rev. A \textbf{49}, 2117--2132 (1994).

\end{thebibliography}

\section{Introduction}

The finite-difference time-domain (FDTD) method is one of the most powerful methods available for solving rigorously Maxwell equations for light propagation in complex settings, such as for self-focusing and four-wave mixing in photonic devices~\cite{taflove2005computational}. However, there are still open fundamental questions and technical challenges associated with the inclusion of atomic-level contributions to that framework~\cite{Varin2014}. These questions and challenges are particularly relevant to modeling sub-wavelength nanophotonic devices that can have only a small number of atoms in one (e.g., a nanofilm) or more dimensions (e.g., a nanorod or a nanoparticle) and where continuum optics theory fails to describe light-matter interaction accurately. In this paper, we develop a three-dimensional (3D) nonlinear atomic polarization model that integrates well in the FDTD method, while retaining the most important atomic response features that usually manifest only through the solution of the quantum mechanical problem~\cite{boyd2008nonlinear}. Ultimately, this model could be used in the microscopic particle-in-cell (MicPIC) framework~\cite{Varin2014,Varin2012,Peltz2012,Peltz2014} to allow \emph{ab initio} modeling of nonlinear optics, where optical materials are represented by ensembles of individual atoms.

Phenomenological and semi-classical models have been proposed, both in frequency and time domain, for the simulation of nonlinear light propagation. But in the FDTD framework,
the inclusion of optical nonlinearities leads to implicit equations that have to be solved itera-
tively~\cite{Goorjian1992,Greene2006,taflove2013advances}. The major drawback with implicit schemes is that they rely on complex implementations and require matrix operations that tend to use a lot of computational resources. 

The nonlinear polarization model presented in this paper has the advantage of being fully explicit while effectively including more physics than models with, say, only an instantaneous Kerr contribution. Moreover, because it is based on an harmonic oscillator model, it can be combined seamlessly with the ordinary differential equations derived from the Sellmeier equation for a particular medium. Ultimately, the saturable oscillator model we propose can be generalized to deal with 3D vectorial propagation of intense laser pulses in anisotropic nonlinear media. 

We recall that there are two possible contributions to the third-order nonlinearity: the electronic Kerr effect, which is considered here, and the vibrational Raman effect~\cite{boyd2008nonlinear}. Besides allowing explicit integration of the Kerr nonlinearity in FDTD, our model also allows to describe the Kerr effect in the high-intensity non-perturbative limit of light-matter interaction, with potential application to modeling laser filamentation~\cite{Berge2007,couaironPR2007}. We emphasize that at these intensities, other effects such as ionization and plasma dynamics become important and have to be considered. Our intention in this paper is not to fully model
laser filamentation but to show that the saturable oscillator model allows to treat Kerr nonlinearity at high intensities numerically in an explicit and stable manner, and in  agreement with quantum mechanical results. The applicability of our model to the Raman nonlinearity will require further investigation.

The paper is divided as follows. In section~\ref{sec:explicit_fdtd}, we show how explicit integration of nonlinear optical response in FDTD is possible. In section~\ref{sec:anharm_osc}, we present the anharmonic oscillator model, an intuitive nonlinear extension to the Lorentz medium model. In section~\ref{sec:quantum_sat}, we review the concepts of nonlinear optics in quantum mechanical two-level systems to explain the origin of the nonlinear response in terms of atomic transition saturation. In section~\ref{sec:sat_harm}, we present the saturable oscillator model and show that it behaves much like a quantum two-level system, up to the 9th harmonic, even in the strong electric fields that characterize laser filamentation. Finally, we draw conclusions in section~\ref{sec:conclusions}.

\section{Explicit nonlinear optics in the FDTD framework\label{sec:explicit_fdtd}}
Based on Yee's seminal work~\cite{yee1966}, modern FDTD methods~\cite{taflove2005computational,taflove2013advances} model light propagation in 
continuous polarizable media by solving the Maxwell's macroscopic curl equations~\cite{stratton1941electromagnetic,jackson1998electromagnetic}. For example, if we assume a nonmagnetic medium ($\mathbf{B} = \mu_0\mathbf{H}$) with a polarization density $\mathbf{P}$, the FDTD technique solves
\begin{subequations}
\begin{align}\label{eq:curl1}
\nabla\times\mathbf{E} &= -\mu_0\frac{\partial \mathbf{H}}{\partial t}\\
\nabla\times\mathbf{H} &= \epsilon_0\frac{\partial \mathbf{E}}{\partial t}+\frac{\partial \mathbf{P}}{\partial t}\label{eq:curl2}
\end{align}
\end{subequations}
using centered finite differences on numerical meshes, staggered in space and time, to achieve second-order accuracy. Finite difference operators and staggering are applied to all electric, magnetic, and electric polarization density components. 

In linear optics, the polarization vector is proportional to the electric field, i.e., $\mathbf{P} = \epsilon_0 \chi^{(1)}\mathbf{E}$, where $\chi^{(1)}$ is the linear susceptibility, a real constant (dispersion is neglected for simplicity). Then, applying the temporal finite difference operator to Eq.~\eqref{eq:curl2},
\begin{align}
\frac{\partial \mathbf{E}}{\partial t}\simeq \frac{\mathbf{E}^{n+1} - \mathbf{E}^{n}}{\Delta t},
\end{align}
one can easily isolate $\mathbf{E}^{n+1}$ and express it in terms of known values of $\mathbf{E}$ and $\mathbf{H}$ at previous times only. Integration is then fully explicit. However, if the polarization is nonlinear in the electric field, for example, if $\mathbf{P} = \epsilon_0 (\chi^{(1)}\mathbf{E} + \chi^{(3)}|\mathbf{E}|^2\mathbf{E})$ like in the presence of instantaneous Kerr nonlinearity~\cite{boyd2008nonlinear}, it is no longer possible to isolate the individual field components. To find $\mathbf{E}^{n+1}$, it is necessary to rely on iterative methods~\cite{Goorjian1992,Greene2006}.

An alternative to the iterative implicit method is to use auxiliary oscillator-type differential equations (AODE). For example, a linear Lorentz medium is modeled by:
\begin{subequations}
\begin{align}\label{eq:lorentz_1}
\frac{\partial \mathbf{P}}{\partial t} &= \mathbf{J}\\
\frac{\partial \mathbf{J}}{\partial t} &= -\omega_0^2\mathbf{P} + \epsilon_0\Omega^2\mathbf{E},
\label{eq:lorentz_2}
\end{align}
\end{subequations}
where $\Omega$ is a coupling frequency, related to the static linear susceptibility $\chi^{(1)}(\omega = 0) = \Omega^2/\omega_0^2$. Using leapfrogged finite differences:
\begin{subequations}
\begin{align}\label{eq:current_fd_a}
\mathbf{J}^{n+1/2} &= \mathbf{J}^{n-1/2}-\omega_0^2\mathbf{P}^{n}\Delta t + \epsilon_0\Omega^2\mathbf{E}^{n}\Delta t\\
\mathbf{P}^{n+1} &= \mathbf{P}^{n} + \mathbf{J}^{n+1/2}\Delta t.
\label{eq:current_fd_b}
\end{align}
\end{subequations}
Then, Eq.~\eqref{eq:curl2} can be written:
\begin{align}\label{eq:curl2_fd}
\mathbf{E}^{n+1} =  \mathbf{E}^{n} + \frac{\Delta t}{\epsilon_0}\left[
\left(\nabla\times\mathbf{H}\right)^{n+1/2} - \mathbf{J}^{n+1/2}\right]
\end{align}
which depends only on values in the past. From here, it is readily seen that fully explicit nonlinear optical material modeling can be achieved in the FDTD framework by replacing the linear driving force $\epsilon_0\Omega^2\mathbf{E}^{n}$ in Eq.~\eqref{eq:current_fd_a} by a nonlinear force $f(\mathbf{E}^{n})$.

The advantages of the AODE approach are the following. First, it naturally introduces a specific time dependence to the polarization and current densities to model dispersion. Second, with a second-order equation for the polarization density $\mathbf{P}$, a current density $\mathbf{J}$ can be defined at the right temporal staggering points for the calculation of the electric field in the future, in a fully explicit manner. Third, because it is based on the harmonic oscillator, it integrates seamlessly to linear models based on the ordinary differential equations derived from the Sellmeier equation for a given medium. Finally, it can be generalized, with the right form of $f(\mathbf{E})$, to take into account tensorial anisotropic response and nonlinear susceptibility of arbitrary order. 

Later, in section~\ref{sec:quantum_sat}, we define $f(\mathbf{E})$ based on the solution for light propagation in a quantum mechanical two-level system. In section~\ref{sec:sat_harm}, we present the saturable oscillator model and show that quantitative agreement can be obtained with the quantum mechanical solution, up to the 9th harmonic. The observation of harmonics with order higher than 10 at high laser intensities ($> 10^{12}\,\mathrm{W/cm}^2$) suggests that $f(\mathbf{E})$ could ultimately be tailored to model high-harmonic generation (HHG). We emphasize the fact that at such high laser intensities field ionization would have to be included to account for the relevant physical processes.

\section{The anharmonic oscillator model\label{sec:anharm_osc}}

The optical response of dielectric media is usually well described in the context of the Lorentz model, where atoms are represented by classical harmonic oscillators~\cite{jackson1998electromagnetic}:
\begin{align}\label{eq:lorentz_ft}
\frac{d^2 \mathbf{P}}{d t^2} + \omega_0^2 \mathbf{P} = \epsilon_0 \Omega^2 \mathbf{E}\quad \overset{\mathrm{FT}}{\Longleftrightarrow}\quad \tilde{\mathbf{P}}(\omega) = \frac{\epsilon_0\Omega^2}{\omega_0^2 - \omega^2}\,\tilde{\mathbf{E}}(\omega).
\end{align}
Note that we neglected damping for simplicity, but it can readily be included by adding $\gamma d\mathbf{P}/dt$ to the left-hand side of the left equation. A natural extension to this model is to add nonlinearity to the restoring force felt by the bound electrons, e.g., for a centrosymmetric material~\cite{boyd2008nonlinear},
\begin{align}\label{eq:anharm_osc}
\frac{d^2 \mathbf{P}}{d t^2} + \omega_0^2 \mathbf{P} - b \left(\mathbf{P}\cdot\mathbf{P}\right)\mathbf{P} = \epsilon_0 \Omega^2 \mathbf{E},
\end{align}
where $b$ is a constant proportional to $\chi^{(3)}$ that modulates the strength of the third-order nonlinearity.

The anharmonic oscillator model presented above is a perturbative approach, where the atomic potential is expanded in powers of the electronic displacement from non-driven equilibrium. It is valid when field excitation is not too strong. Its fully explicit integration in the FDTD/particle-in-cell (PIC) frameworks was demonstrated recently by Gordon, Helle, and Pe\~nano~\cite{Gordon2013}. However, as the authors reported, it tends to be unstable. 

In the next two sections, we show how the Lorentz model of the atom can be extended to deal with nonlinear optics, not by adding inherently unstable nonlinear restoring force contributions, but by using the proper nonlinear driving force. The model we propose also supports fully explicit integration in the FDTD/PIC frameworks and can be seen as a drop-in replacement for the method proposed by Gordon, Helle, and Pe\~nano~\cite{Gordon2013}.

\section{Nonlinear saturation in a two-level system\label{sec:quantum_sat}}
The quantum mechanical two-level model provides a very good approximation to a more complete theory of nonlinear light-matter interaction~\cite{boyd2008nonlinear,allen1975optical}. Solution of the two-level optical Bloch equations in the under-resonant limit yields the polarization density of an ensemble of two-level atoms subjected to an electric field as~\cite{boyd2008nonlinear}
\begin{align}\label{eq:quantum_sol}
P\simeq \frac{\epsilon_0\widetilde{\chi}_0}{1+\Delta^2\tau^2 + |E|^2/|E_s|^2}\,E,
\end{align}
where $\widetilde{\chi}_0$ is a complex-valued $\omega$-dependent weak-field susceptibility, $\Delta = \omega - \omega_0$ is the angular frequency relative to the resonance (transition) frequency $\omega_0$, $\tau$ is the induced dipole moment decay time, and $E_s$ is the line-center saturation field strength.
If the denominator is expanded in a power series in terms of $|E|^2/|E_s|^2$, it gives
\begin{subequations}
\begin{align}
P&\simeq\epsilon_0\widetilde{\chi}_0\left(\frac{1}{1+\Delta^2\tau^2} - \frac{1}{\left[1+\Delta^2\tau^2\right]^2}\frac{|E|^2}{|E_s|^2}+\frac{1}{\left[1+\Delta^2\tau^2\right]^3}\frac{|E|^4}{|E_s|^4}-\ldots\right)\,E\\
&\equiv \epsilon_0\left(\chi^{(1)} - \chi^{(3)}|E|^2 + \chi^{(5)}|E|^4 -\ldots \right)E.
\label{eq:quantum_sol_expand}
\end{align}
\end{subequations}
We emphasize that this expansion is valid in the weak field limit only ($|E|^2\ll|E_s|^2$). The most important feature of Eq.~\eqref{eq:quantum_sol}---the saturation of the susceptibility in the presence of strong fields---is equally lost in Eq.~\eqref{eq:quantum_sol_expand} and the anharmonic oscillator model. 

Finally, quantum mechanics suggests that a driving function of the form 
\begin{align}\label{eq:f_e_centro}
f(E) = \frac{E}{1 + E^2/E_s^2}
\end{align}
introduces a nonlinear polarization that expands, in the weak field limit, to the power series that is usually used to model centrosymmetric nonlinear media~\cite{boyd2008nonlinear}. It is straightforward to see that the expansion of
\begin{align}\label{eq:f_e_noncentro}
f(E) = \frac{E}{1 + E/E_s}
\end{align}
has even and odd orders in $E$ and can potentially be used to model noncentrosymmetric nonlinear media. We emphasize that for an accurate description of optical saturation in atoms higher energy levels need to be taken into account. This will change the value of the saturation field strength $E_s$, but not the structure of the expression.

We will see in the next section that replacing $\mathbf{E}$ by $f(\mathbf{E})$ in Eq.~\eqref{eq:lorentz_2} is actually a very good approximation to solving the quantum mechanical two-level equations, with the advantage that it exposes naturally the vectorial nature of the problem.

\section{The saturable oscillator model\label{sec:sat_harm}}

The saturable harmonic oscillator model is characterized by the following equation:
\begin{align}\label{eq:sat_harm_model}
\frac{d^2 \mathbf{P}}{d t^2} + \omega_0^2 \mathbf{P} = \frac{\epsilon_0 \Omega_s^2}{1+ |\mathbf{E}|^2/E_s^2}\,  \mathbf{E}.
\end{align}
Actually, it is the Lorentz model [see Eq.~\eqref{eq:lorentz_ft}] where the driving force is modulated by a saturation function $1/(1+ |\mathbf{E}|^2/E_s^2)$ whose expression is inspired by the quantum mechanical solution presented in section~\ref{sec:quantum_sat}. The angular frequency $\Omega_s$ and saturation field strength $E_s$ are fitting parameters that we will use below to match a particular medium response. Equation~\eqref{eq:sat_harm_model} can be solved with leapfrogged finite differences, as we have shown earlier at Eqs.~\eqref{eq:current_fd_a}-\eqref{eq:current_fd_b}, i.e.,
\begin{subequations}
\begin{align}\label{eq:sat_harm_model_fd}
\mathbf{J}^{n+1/2} &= \mathbf{J}^{n-1/2}-\omega_0^2\mathbf{P}^{n}\Delta t + \frac{\epsilon_0 \Omega_s^2}{1+ |\mathbf{E}^n|^2/E_s^2}\,\mathbf{E}^{n}\Delta t\\
\mathbf{P}^{n+1} &= \mathbf{P}^{n} + \mathbf{J}^{n+1/2}\Delta t.
\end{align}
\end{subequations}
We have seen in section~\ref{sec:explicit_fdtd} that this particular form allows fully explicit integration of the electromagnetic field components with the FDTD method. 

In the strong field limit ($|\mathbf{E}^n| \lesssim E_s$), the driving force starts saturating, which introduces nonlinearities. In the weak field limit ($|\mathbf{E}^n|\ll E_s$), after the power series expansion of the saturation term, one gets an anharmonic oscillator equation similar to Eq.~\eqref{eq:anharm_osc}. 

Next, we test the saturable oscillator model numerically with parameters that characterize ultrashort pulse filamentation in air. In particular, in section~\ref{subsec:parameters}, we take advantage of the fact that the leading term of the weak field expansion is the harmonic oscillator to obtain a value for $\Omega_s$ from the Sellmeier equation for air. In turn, the critical power for self-focusing allows the definition of $\chi^{(3)}$ and $E_s$. Ultimately, in section~\ref{subsec:results}, we compare the saturable oscillator model against those presented in the previous sections.

\subsection{Parameters for nonlinear propagation in air\label{subsec:parameters}}

Laser filamentation is a spectacular manifestation of the richness of non-perturbative nonlinear optics. Besides technical applications for long-distance propagation of intense laser beams in the atmosphere, harmonic generation, and supercontinuum emission, there remain fundamental questions~\cite{couaironPR2007,Berge2007}. There is, in particular, the debated question about the contribution of nonlinear optical saturation to the stopping of self-focusing, which is usually attributed to plasma generation alone~\cite{Berge2007}. Recent quantum computations performed in the context of laser filamentation in gases display evidences that only relying on instantaneous higher-order Kerr nonlinearities while ignoring plasma generation is not possible at intensities of the order of $10^{13}\,\mathrm{W/cm}^2$ and beyond~\cite{Volkova2011,Kohler2013,Spott2014}. However, experimental observations show that during filamentation the medium is only weakly ionized, i.e., most electrons remain bound. These bound electrons contribute to the Kerr nonlinearity, even for the high intensities that characterize the filament core, accepted to be in the $10^{13}-10^{14}\,\mathrm{W/cm}^2$ range~\cite{Berge2007,couaironPR2007}. The FDTD method, combined with the saturable oscillator model and proper ionization/plasma models~\cite{couaironPR2007}, is a unique tool to study those questions in a setting that takes into account the temporal, full vectorial, and 3D nature of the phenomenon.

For the following numerical tests, we applied the saturable oscillator model to air, because of the importance of that medium in laser filamentation~\cite{couaironPR2007,Berge2007}. Our intention is not to fully model laser filamentation but to show that our simple model allows to treat Kerr nonlinearity at high intensities in a numerically stable manner and in  agreement with quantum mechanical results. Based on the Sellmeier equation presented in~\cite{Ciddor96}, the linear response of air in the infrared range of the electromagnetic spectrum is accurately modeled by the following susceptibility:
\begin{align}
\chi^{(1)}(\omega) = \left(\frac{\Omega_1^2}{\omega_1^{2} - \omega^2}\right) + \left(\frac{\Omega_2^2}{\omega_2^{2} - \omega^2}\right),
\end{align}\\[-0.45cm]
where 
\begin{subequations}
\begin{align}
\Omega_1 &= 6.411\times 10^{14}\,\mathrm{rad/s} \quad &\omega_1 = 2.906\times 10^{16}\,\mathrm{rad/s}\\
\Omega_2 &= 1.092\times 10^{14}\,\mathrm{rad/s}\quad &\omega_2 = 1.427\times 10^{16}\,\mathrm{rad/s}
\end{align}
\end{subequations}
From the critical power for self-focusing of a Gaussian beam in air at the wavelength of 800~nm ($\sim  3.2\,\mathrm{GW}$)~\cite{couaironPR2007}, we inferred $n_2\simeq 3\times 10^{-23}\,\mathrm{m}^2/\mathrm{W}$ and $\chi^{(3)}\simeq 1.062\times 10^{-25}\, \mathrm{m}^2/\mathrm{W}\Omega$. Then we define the weak field nonlinear polarization density of air as:
\begin{align}\label{eq:P_series_air}
\mathbf{P}(\omega) = \epsilon_0\left(\chi^{(1)}  - \chi^{(3)} |\mathbf{E}|^2\right)\mathbf{E}.
\end{align}
If, for simplicity, we compare only the first band ($\omega_1$) near $\omega = 0$ with the static solution of Eq.~\eqref{eq:sat_harm_model}
\begin{align}\label{eq:sat_harm_model_static}
\mathbf{P} = \frac{\epsilon_0 \Omega_s^2/\omega_0^2}{1+ |\mathbf{E}|^2/E_s^2}\,  \mathbf{E} 
\simeq \epsilon_0 \left(\frac{\Omega_s^2}{\omega_0^2} - \frac{\Omega_s^2}{\omega_0^2}\frac{|\mathbf{E}|^2}{E_s^2}\right)\,  \mathbf{E},
\end{align}
we get immediately the following correspondence:
\begin{align}\label{eq:correspondence}
\Omega_1 = \Omega_s\qquad \omega_1 = \omega_0 \qquad \chi^{(3)} = \frac{\Omega_s^2}{\omega_0^2E_s^2}
\end{align}
With the parameters defined so far, it gives $E_s \simeq 6.77\times 10^{10}\,\mathrm{V/m}$, which corresponds to the saturation intensity $I_s = E_s^2/2\eta_0 \simeq 6\times 10^{14}\,\mathrm{W/cm}^2$, where $\eta_0$ is the characteristic impedance of vacuum.

As seen at Eq.~\eqref{eq:correspondence}, the saturable oscillator model gives the freedom to chose the linear and nonlinear refractive index independently, through the parameters $\Omega_s$, $\omega_0$, and $E_s$. Although we applied the model to a single resonance of air, generalization to multiple frequencies is straightforward and done by using a specific saturable oscillator for each term of the Sellmeier equation associated with a given material.

\subsection{Numerical results\label{subsec:results}}
We compared the saturable oscillator model against the anharmonic oscillator model [Eq.~\eqref{eq:anharm_osc}], the quantum mechanical two-level model~\cite{boyd2008nonlinear,allen1975optical}, and the instantaneous Kerr model [Eq.~\eqref{eq:P_series_air}]. In particular for the quantum mechanical two-level model, we integrated numerically Eqs.~(6.5.6) and (6.5.8) found in~\cite{boyd2008nonlinear}, along with the dipole moment defined at Eq.~(6.5.31) of the same reference. In all cases, the driving electric field was that of a Gaussian pulse 
\begin{align}
E(t) = E_0\cos(\omega_l t)\exp(-t^2/T^2),
\end{align}
with angular frequency $\omega_l = 2\pi c/\lambda_l$ ($\lambda_l = 800\,\mathrm{nm}$) and pulse duration $T=10\,\mathrm{fs}$. The field amplitude is defined in terms of the laser intensity $I = E_0^2/2\eta_0$. For all models we used the parameters defined in section~\ref{subsec:parameters}.

\begin{figure}[h]
\begin{center}
\includegraphics[width=\textwidth]{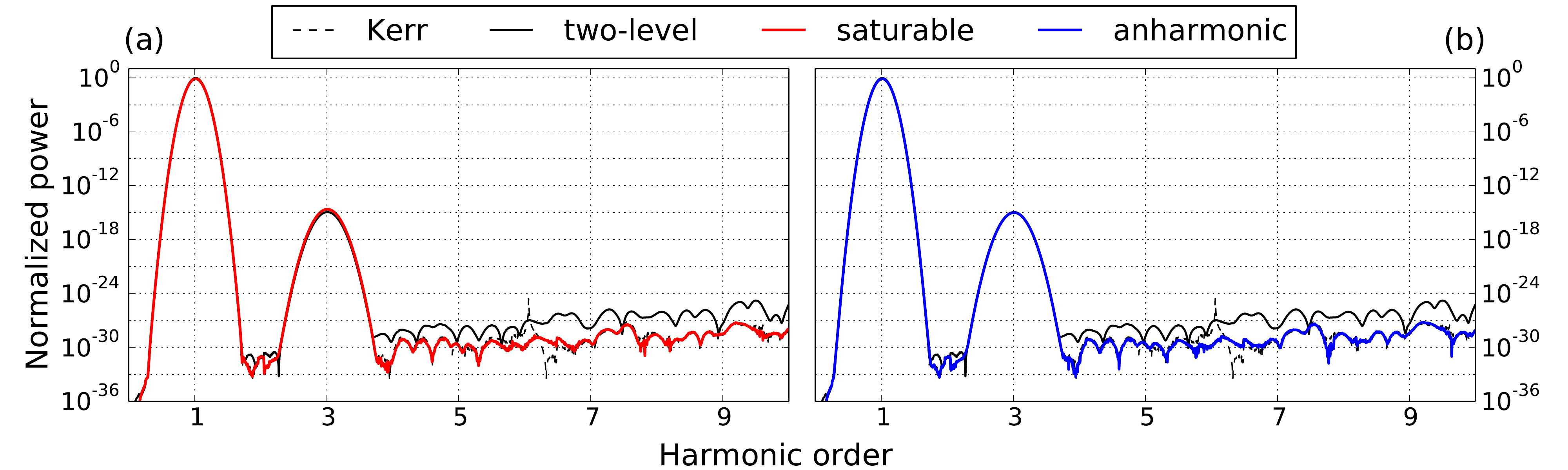}
\caption{Comparison between the saturable oscillator model [Eq.~\eqref{eq:sat_harm_model}] and the other three nonlinear polarization models described in this paper: the anharmonic oscillator model [Eq.~\eqref{eq:anharm_osc}], the quantum mechanical two-level model~\cite{boyd2008nonlinear,allen1975optical}, and the instantaneous Kerr model [Eq.~\eqref{eq:P_series_air}]. The curves represent the power emitted as a function of the normalized angular frequency (harmonic order). Power spectra are normalized by the value at the central laser frequency ($\omega = \omega_l$) of the power spectrum obtained with the instantaneous Kerr model. For direct visual comparison are shown the instantaneous Kerr and two-level results against (a) the saturable oscillator model with $I_s = 3\times 10^{14}\,\mathrm{W/cm^2}$ and (b) the anharmonic oscillator model. For this graph, laser intensity is  $I = 10^{7}\,\mathrm{W/cm^2}$.}
\label{fig:1}
\end{center}
\end{figure}

In particular, we compared the harmonic power spectrum generated by the time-dependent nonlinear polarization density $\mathbf{P}(t)$ by inserting the Fourier transform $\mathbf{P}(\omega)$ into the Larmor radiation formula~\cite{stratton1941electromagnetic}
\begin{align}
P_{rad}(\omega) = \frac{V^2\omega^4}{6\pi\epsilon_0c^3}|\mathbf{P}(\omega)|^2.
\end{align}
All curves were normalized by the value at the peak of the linear response ($\omega/\omega_l = 1$) of the instantaneous Kerr model. Individual model parameters were optimized to reproduce the instantaneous Kerr curve in the perturbation limit where only the first and third harmonics are present, shown in figure~\ref{fig:1}. In this regime all models agree very well, as expected.

\begin{figure}
\begin{center}
\includegraphics[width=\textwidth]{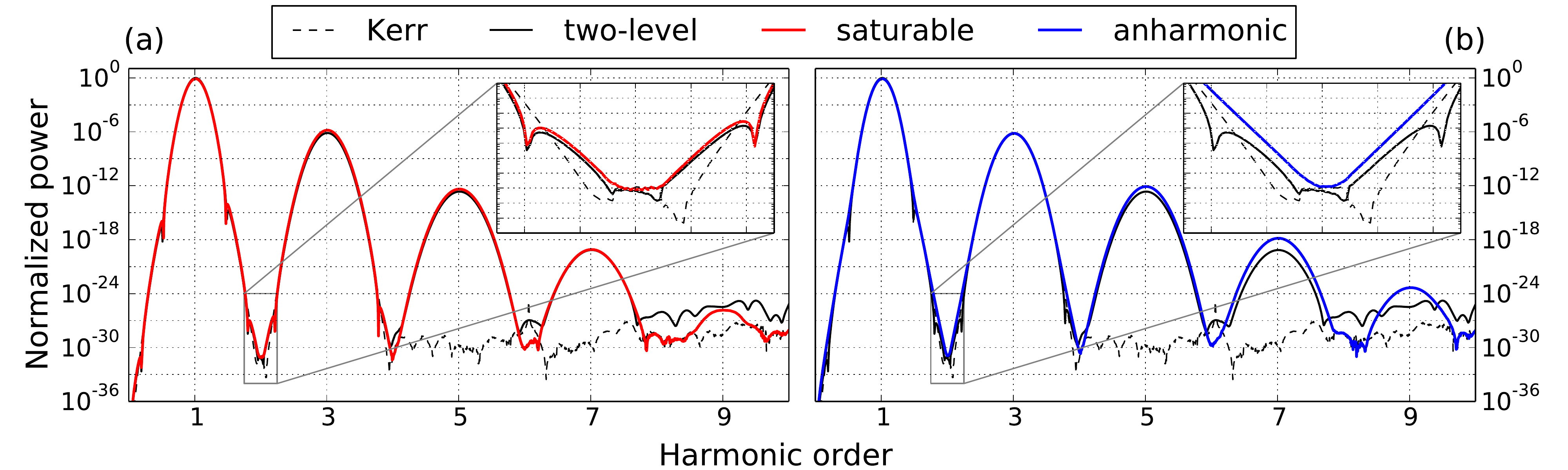}
\caption{Comparison between all four models for $I = 2.5\times 10^{11}\,\mathrm{W/cm^2}$ (see also caption of figure~\ref{fig:1}). In insets, a zoomed view of the curves show that the saturable oscillator reproduces fairly well the phase relationship between frequency components of the quantum mechanical solution, while the anharmonic oscillator does not. Also, the anharmonic oscillator tends to overestimate the 7th and 9th harmonics by a few orders of magnitude.}
\label{fig:2}
\end{center}
\end{figure}

At the beginning of the laser filamentation process, a Gaussian beam propagating in a Kerr medium with a power above the critical power will start self-focusing under the action of the intensity-dependent refractive index~\cite{boyd2008nonlinear}. During that phase, the peak intensity increases rapidly to reach $10^{11}-10^{12}\,\mathrm{W/cm^2}$~\cite{couaironPR2007,Berge2007}. Results obtained in this regime are shown in figure~\ref{fig:2}. From the numerical results it appears that the 5th harmonic is already stronger than the 3rd harmonic in the lower intensity example of figure~\ref{fig:1}, suggesting that it could have a similar relative influence. Overall, the two-level, saturable oscillator, and anharmonic oscillator models agree well, although the anharmonic oscillator overestimates the 7th and 9th harmonics by a few orders of magnitude and fails at reproducing  the dips of destructive interference predicted by both the two-level and saturable oscillator solutions (some are also predicted by the Kerr model, but are not shown).

\begin{figure}
\begin{center}
\includegraphics[width=\textwidth]{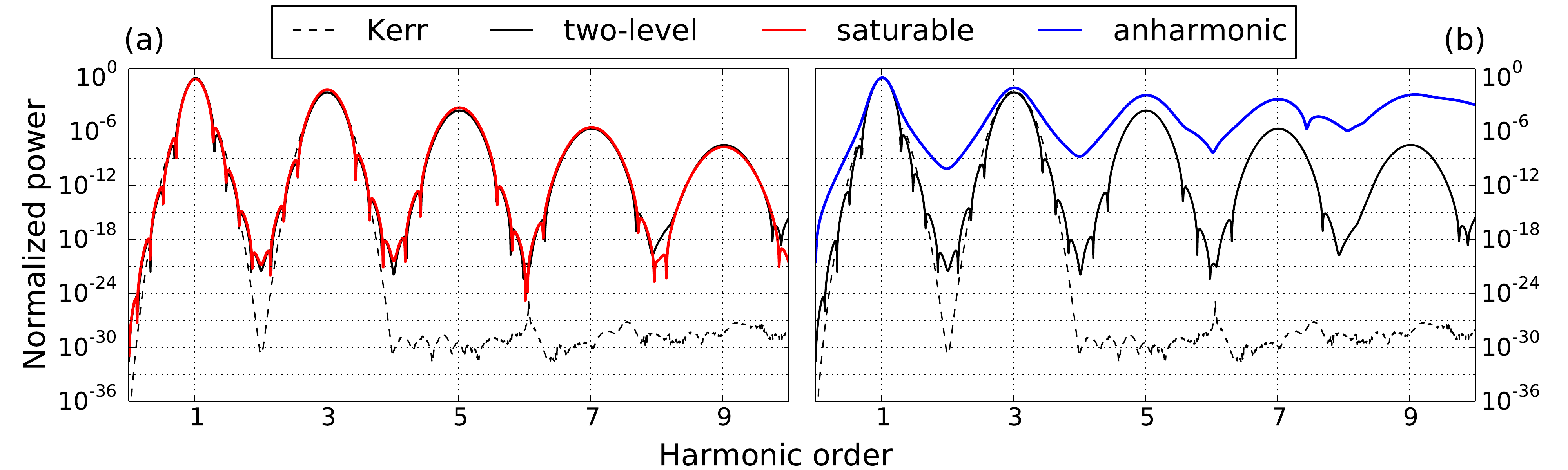}
\caption{Comparison between all four models for $I = 5\times 10^{13}\,\mathrm{W/cm^2}$, a laser intensity comparable to the peak intensity inside a light filament in air~\cite{couaironPR2007,Berge2007} (see also caption of figure~\ref{fig:1}). Whereas the Kerr, two-level, and saturable oscillator models agree, the aharmonic oscillator fails at reproducing the dynamics of the two-level system. With an intensity just a bit higher, at $I = 6.7\times 10^{13}\,\mathrm{W/cm^2}$, the numerical integration of the anharmonic oscillator does not converge anymore.}
\label{fig:3}
\end{center}
\end{figure}

In the widely accepted picture of laser filamentation in air, self-focusing is slowed down and stopped by a defocusing plasma formed due to multiphoton ionization of $\mathrm{O}_2$ molecules, when the intensity reaches $10^{12}-10^{13}\,\mathrm{W/cm^2}$~\cite{couaironPR2007,Berge2007}. The so-called filament emerges from the balance between nonlinear self-focusing and plasma defocusing. Peak intensity is then in the $10^{13}-10^{14}\,\mathrm{W/cm^2}$ range. Atomic polarization in that regime is shown in figure~\ref{fig:3}, for the four models. Let aside the fact that the Kerr model gives only first and third order contributions, the Kerr, two-level, and saturable oscillator models agree very well, whereas the anharmonic oscillator model is completely off by several orders of magnitude at the 5th, 7th, and 9th harmonics. One can see a manifestation of the instability reported by Gordon, Helle, and Pe\~nano~\cite{Gordon2013} in figure~\ref{fig:3}(b), where the 9th harmonic peak is overshooting the 7th and the 5th. Ultimately, at $\sim 6.7 \times 10^{13}\,\mathrm{W/cm^2}$ the 3rd harmonic overshoots the linear peak and the numerical integration of the anharmonic oscillator does not converge anymore, no matter how small the time step is. We emphasize that the anharmonic oscillator instability will show at much lower intensities when modeling solid-density materials because the nonlinear susceptibility is much higher.

\begin{figure}
\begin{center}
\includegraphics[width=0.625\textwidth]{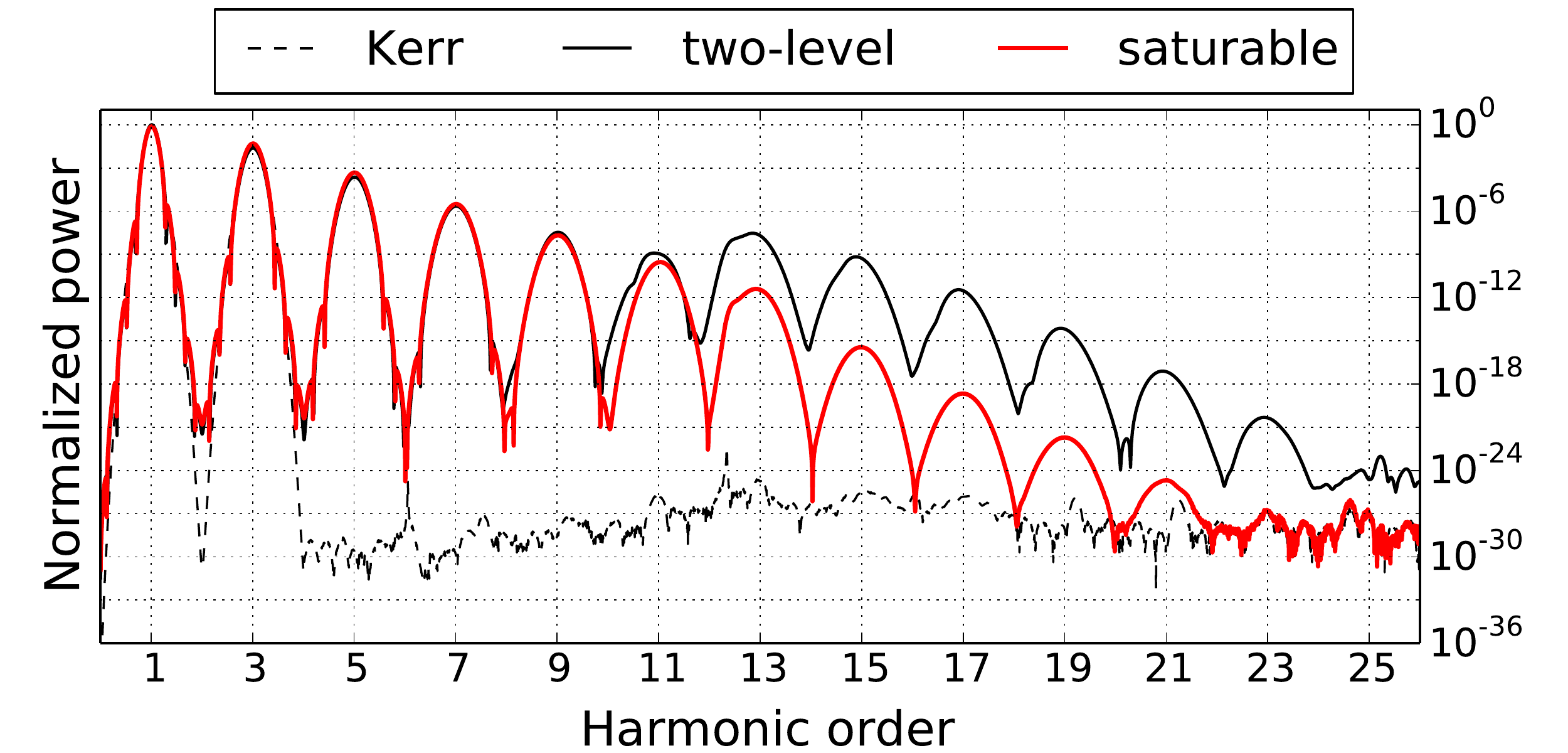}
\caption{The expanded view of figure~\ref{fig:3}(a) shows that agreement between the two-level and saturable oscillator models is up to the 9th harmonic. The observation of harmonic orders beyond 10 suggests that the nonlinear driving force of the saturable oscillator could be tailored to model HHG.}
\label{fig:4}
\end{center}\vspace{-0.45cm}
\end{figure}

An expanded view of figure~\ref{fig:3}(a) is given in figure~\ref{fig:4}. It is readily observed that the agreement between the quantum mechanical two-level model and the classical saturable oscillator model is almost perfect up to the 9th harmonic, but not beyond. Finally, we emphasize that HHG in gases is commonly performed in the $10^{14}-10^{15}\,\mathrm{W/cm^2}$ intensity range, where the atomic polarization model presented here needs to be complemented by the ionization-recollision process~\cite{corkumPRL1993,LewensteinPRA1994}.

\section{Conclusions\label{sec:conclusions}}
In this paper, we have demonstrated that the saturable oscillator model behaves effectively much like a quantum mechanical two-level system. The advantage of the saturable oscillator over the two-level model is the possibility to extend---in a simple and intuitive way---the Sellmeier-harmonic oscillator formalism to include damping, restoring force, and driving force coupling in a tensorial form to allow effective modeling of the anisotropic nonlinear response and excitation decay of specific crystalline structures and molecules. We have shown that numerical integration can be fully explicit, allowing straightforward integration to the FDTD, PIC, and MicPIC frameworks with, ultimately, the possibility to model nonlinear light propagation with atomic scale resolution. In particular, we have shown that the saturable oscillator model gives accurate results up to the 9th harmonic, for laser intensities relevant to the process of laser filamentation in air.
\vspace{2pt}

\end{document}